\documentclass[prc,twocolumn,superscriptaddress,superscriptfootnote,nofootinbib]{revtex4}
\usepackage{amssymb}

\usepackage{amsmath,amsfonts,amssymb,epsfig,color}

\begin{document}
\title{Microscopic shell-model counterpart of the Bohr-Mottelson model}
\author{H. G. Ganev}
\affiliation{Joint Institute for Nuclear Research, Dubna, Russia
and\\
Institute of Nuclear Research and Nuclear Energy, Bulgarian Academy
of Sciences, Sofia, Bulgaria}

\setcounter{MaxMatrixCols}{10}

\begin{abstract}
In the present paper we demonstrate that there exists a fully
microscopic shell-model counterpart of the Bohr-Mottelson model by
embedding the latter in the microscopic shell-model theory of atomic
nucleus within the framework of the recently proposed fully
microscopic proton-neutron symplectic model (PNSM). For this
purpose,  another shell-model coupling scheme of the PNSM is
considered in which the basis states are classified by the algebraic
structure $SU(1,1) \otimes SO(6)$. It is shown that the
configuration space of the PNSM contains a six-dimensional subspace
that is closely related to the configuration space of the
generalized quadrupole-monopole Bohr-Mottelson model and its
dynamics splits into radial and orbital motions. The group $SO(6)$
acting in this space, in contrast, e.g., to popular IBM, contains an
$SU(3)$ subgroup which allows to introduce microscopic shell-model
counterparts of the exactly solvable limits of the Bohr-Mottelson
model that closely parallel the relationship of the original
Wilets-Jean and rotor models. The Wilets-Jean-type dynamics in the
present approach, in contrast to the original collective model
formulation, is governed by the microscopic shell-model intrinsic
structure of the symplectic bandhead which defines the relevant
Pauli allowed $SO(6)$, and hence $SU(3)$, subrepresentations. The
original Wilets-Jean dynamics of the generalized Bohr-Mottelson
model is recovered for the case of closed-shell nuclei, for which
the symplectic bandhead structure is trivially reduced to the scalar
or equivalent to it irreducible representation.
\end{abstract}
\maketitle

PACS number(s): {21.60.Fw, 21.60.Ev}

\section{Introduction}

Majority of atomic nuclei exhibit collective behavior in their
spectra, which is primarily represented by the nuclear rotation.
Indeed, the rotational states are prevalent in atomic nuclei. One of
the fundamental models of nuclear structure is the Bohr-Mottelson
(BM) collective model \cite{BM}, which has demonstrated that nuclear
collective motion can be described by considering only few
macroscopic collective degrees of freedom. Conceptually, this model
has an invaluable impact on the understanding of nuclear collective
motion and the development of nuclear structure models. Moreover,
for heavy nuclei, the BM collective model provides the basic
concepts and language in terms of which the nuclear collective
phenomena are described.

It is well known that the BM model has three algebraically solvable
limits \cite{BM}: the harmonic vibrator model, the $\gamma$-unstable
Wilits-Jean (WJ) model \cite{WJ}, and the rigid-rotor model
\cite{DF,rot3}. These solvable submodels provide approximate
descriptions of a subset of nuclear collective states. The
group-theory of the BM model was given in \cite{GT-BM}, where the
conventional $U(5) \supset SO(5)$ harmonic-vibrational basis is
exploited. This basis is actually used in practical applications of
the BM model by the Frankfurt school \cite{Frankfurt}. The physics
and mathematics of the exactly solvable limits of the Bohr-Mottelson
model, as well as their relationships, are presented in an
exhaustive manner in a recent book \cite{RW}.

The BM model has been formulated also in algebraic terms by means of
different spectrum generating algebras (SGA) and dynamical groups.
The position and momentum coordinates of the BM model, for example,
close the Lie algebra of Heisenberg-Weyl group
$HW(5)=\{\alpha_{\mu},\pi^{\nu},I\}$. It is too small to contain
useful subgroup chains with which to classify basis states, but it
provides the basic building blocks from which numerous dynamical
groups and spectrum generating algebras can be constructed. Among
them, the following two are important for our present
considerations, namely $[HW(5)]U(5)$ and $SU(1,1) \otimes SO(5)$
dynamical groups \cite{RW}. The latter turns out to be a very
efficient one and allowed the formulation of a powerful version of
the BM model, called Algebraic Collective Model (ACM)
\cite{Rowe-ACM1,Rowe-ACM2}. For spherical nuclei, the $SU(1,1)
\otimes SO(5)$ basis of the ACM reduces to that of the
five-dimensional harmonic oscillator and is given by the harmonic
series of $SU(1,1)$ irreps. For deformed nuclei, the modified
oscillator series \cite{Rowe-Euclidean} of $SU(1,1)$ irreps give
much more rapidly convergent results. The three classical BM
submodels have been expressed in terms of the $[HW(5)]U(5)$
dynamical group and its subgroups. Thus, the three dynamical groups
$U(5)$, $[R^{5}]SO(5)$ and $[R^{5}]SO(3)$ have been shown to
correspond to the spherical vibrator, Wilets-Jean $\gamma$-unstable
and rigid-rotor limits of the BM model, respectively
\cite{RW,Rowe-ACM1}. The irreducible representations of the
$[R^{5}]SO(5)$ group, as first shown by Elliott \emph{et al.}
\cite{Elliott86b}, are characterized by rigid value of
$\beta=\beta_{0}$, i.e. by a sharp value of the quadrupole moment.
In turn, the rigid-rotor model irreps are characterized by both
$\beta$ and $\gamma$ rigid values $\beta_{0}$ and $\gamma_{0}$,
respectively.

There is a close correspondence of the physics of the BM model to
that of the Interacting Boson Model (IBM) \cite{IBM}, between which
many relationships have been established \cite{RW,RoweThiamova05}
using the known fact that two finite Hilbert spaces of equal
dimension are isomorphic to each other. In this regard, the IBM
achieves a reduction of the non-compact BM state space to a finite
dimensional space by compactifying the algebraic structure
$[HW(5)]U(5)$ of the collective model to the $U(6)$ group of
six-dimensional harmonic oscillator. It was also pointed out that
the IBM could be considered as an algebraic approximation to the
Bohr-Mottelson collective model \cite{Elliott85}. As will be shown
further, although the WJ and rigid-rotor models have algebraic
structures and their states are characterized by certain dynamical
group chains, they are not particularly useful in the construction
of square-integrable wave functions due to their delta function
nature. This problem was circumvented in the ACM \cite{Rowe-ACM1,RW}
by relaxing the rigidity of WJ model by replacing its dynamical
group $[R^{5}]SO(5)$ with $SU(1,1) \otimes SO(5)$ one, which results
in a more physical collective model. In fact, the ACM is a
computationally tractable version of the BM collective model, which
make use of the $\beta$ wave functions given analytically by the
softened-$\beta$ version of the WJ model, initially considered by
Elliott \emph{et al.} \cite{Elliott86a,Elliott86b}. This allows the
$\beta$-rigid and $\gamma$-rigid limits to be approached in a
continuous way with increasingly narrow but square-integrable
$\beta$ and $\gamma$ wave functions.

The IBM also has three exactly solvable dynamical symmetry limits
that correspond to similar dynamical symmetries of the BM model. A
trivial relationship between the two models is obtained for the
five-dimensional spherical vibrator submodel of BM model which
corresponds to the $U(5)$ limit of the IBM. The $\beta$-rigid but
$\gamma$-unstable WJ model has firstly been shown by J.
Meyer-ter-Vehn \cite{Meyer-ter-Vehn79} to correspond to the $SO(6)
\supset SO(5) \supset SO(3)$ dynamical symmetry limit of the IBM.
Further, it turns out that there is \emph{no} analogue in the IBM of
the $\beta$-rigid and $\gamma$-rigid rotor submodel of the BM model,
which is obviously a submodel of the $\beta$-rigid but
$\gamma$-unstable WJ model since $[R^{5}]SO(3) \subset
[R^{5}]SO(5)$. This is because in IBM the $SU(3)$ dynamical group,
associated with the rotational states, is not a subgroup of $SO(6)$.
Hence, the rotor-like states in the $SU(3)$ limit of the IBM are
\emph{not} related to those of its $SO(6)$ limit in a way that
parallels the relationship between the rigid-rotor and WJ states in
the BM model. This fact was stressed in Ref.\cite{RoweThiamova05}.
In this regard, it is the purpose of present work to demonstrate
that there exists a microscopic many-particle shell-model
counterpart of the Bohr-Mottelson model whose exactly solvable
limits have a relationship that closely resembles the one between
the original WJ and rigid-rotor submodels.

From another side, it is known that a limitation of the BM model is
that it has irrotational-flow moments of inertia which are much
smaller than those needed to describe the low-energy rotational
states of deformed nuclei. Usually this problem is resolved by
treating the moments of inertia as free parameters which are fitted
to the experimental spectra of nuclei. Further, in its standard
formulation, the BM model can not be naturally related to the
microscopic shell-model theory of nucleus. In particular, the
vectors in the BM model which define the states of a
quantum-mechanical liquid drop cannot be identified with the wave
functions in the many-particle Hilbert space of $A$ nucleon
antisymmetric states. The problem of incorporating the BM model into
the microscopic theory of the nucleus and its importance for nuclear
structure physics have been realized long time ago. The solution of
this problem was given through the algebraic approach. It was shown
(see, e.g. \cite{stretched,Rowe96}) that the collective model of
Bohr and Mottelson admits a microscopic realization first by
augmenting it by vorticity degrees of freedom, important for the
appearance of low-lying collective states, and second by making it
compatible with the composite many-fermion structure of the nucleus.
The result is the one-component $Sp(6,R)$\footnote{Throughout the
present work, we will use the notation Sp(2n,$R$) for the group of
linear canonical transformations in $2n$-dimensional phase space.
Some authors denote the Sp(2n,$R$) group by Sp(n,$R$).} symplectic
model \cite{RR1} of nuclear collective motion, sometimes called a
microscopic collective model, which is a submodel of the nuclear
shell model. The $Sp(6,R)$ model of nuclear rotations, among its
submodels, contains the rigid-rotor model \cite{rot3} and the
Elliott's $SU(3)$ shell model of collective rotations
\cite{Elliott58}. The presence of vorticity in the $Sp(6,R)$ model
results in a complete range of possible collective flows from
irrotational-flow (zero vorticity) to rigid rotations. This is of
significant importance as well as the fact that the vortex-spin
degrees of freedom are responsible for the appearance of low-lying
collective states \cite{stretched,Rowe96}. However, the microscopic
collective model, which is just a microscopic version of the BM
model augmented by the vortex spin degrees of freedom and compatible
with the many-particle nucleon structure of nucleus, does not
contain an $SO(6)$ algebraic structure that could allow to establish
a close relationship to the WJ model in a manner similar to that of
IBM. In this way, the findings of the present work in embedding the
generalized Bohr-Mottelson model in the microscopic shell-model
theory of the nucleus have a more natural interpretation of the
underlying BM quadrupole-monopole collective dynamics than in the
microscopic realization of the BM collective model provided by the
one-component $Sp(6,R)$ symplectic model.

Recently, a fully microscopic proton-neutron symplectic model (PNSM)
of nuclear collective motion with $Sp(12,R)$ dynamical algebra was
proposed by considering the symplectic geometry and possible
collective flows in the two-component many-particle nuclear system
\cite{cdf}. Through its more general motion group $GL(6,R) \subset
Sp(12,R)$, which allows for the separate treatment of the collective
dynamics of proton and neutron subsystems as well as the combined
proton-neutron collective excitations, the PNSM generalizes the
$Sp(6,R)$ model for the case of two-component proton-neutron
many-particle nuclear systems. This can be easily understood by the
embedding $Sp(6,R) \subset Sp(12,R)$. The configuration space of the
PNSM is isomorphic to the coset space $GL(6,R)/SO(6)$ and is spanned
by the commuting quadrupole momentum observables, i.e.
$\mathbb{R}^{21} = \{Q_{ij}(\alpha,\beta)\}$ (cf. Eq.(\ref{Sp12a})).
This configuration space contains a six-dimensional subspace of the
combined proton-neutron collective dynamics $\mathbb{R}^{6} \subset
\mathbb{R}^{21}$ that is closely related to the configuration space
of the generalized BM model, in which the monopole degrees of
freedom are also included. The group $SO(6)$ acting in
$\mathbb{R}^{6} \subset \mathbb{R}^{21}$ contains an $SU(3)$
subgroup, as will be demonstrated further, and allows us to
introduce a microscopic shell-model counterpart of the BM model,
whose two exactly solvable limits closely parallel the relationship
of the original WJ and rotor models. For this purpose, we first
shortly consider the classical versions of the original BM
submodels. Then we consider the reformulation of the BM limits in
algebraic terms and consider their relation to the IBM, which will
allow a better understanding of the close relationships between the
WJ and rigid-rotor models and their parallel construction within the
framework of the PNSM. This reveals further dynamical content of the
latter, not considered before.

The present paper deals with another shell-model coupling scheme
within the framework of the microscopic proton-neutron
symplectic-based shell-model approach. In principle, this coupling
scheme provides an alternative basis for the shell-model
diagonalization of an arbitrary collective Hamiltonian, which could
also be expressed as a polynomial in the many-particle position and
momentum coordinates of the two-component proton-neutron nuclear
systems. This will extend the applicability of the PNSM in
describing the collective properties in various nuclei. It will be
interesting also to compare the results obtained within the two
shell-model coupling schemes of the PNSM introduced so far.

\section{Bohr-Mottelson submodels}

The configuration space of the BM model is $\mathbb{R}^{5}$, i.e. it
has the geometry of Euclidean space. The volume element in spherical
coordinates is given by
\begin{equation}
dV = \beta^{4}d\beta sin3\gamma d\gamma d\Omega, \notag
\end{equation}
where $d\Omega$ is the $SO(3)$ volume element. The Laplacian
operator is
\begin{equation}
\nabla^{2} =
\frac{1}{\beta^{4}}\frac{\partial}{\partial\beta}\beta^{4}\frac{\partial}{\partial\beta}
-\frac{\Lambda^{2}}{\beta^{2}}, \label{Laplacian}
\end{equation}
where the $SO(5)$ Casimir operator $\Lambda^{2}$ is expressed in
terms of the intrinsic $SO(3)$ angular momentum operators
$\{\overline{L}_{k}\}$ as
\begin{equation}
\Lambda^{2} = -
\frac{1}{sin3\gamma}\frac{\partial}{\partial\gamma}sin3\gamma\frac{\partial}{\partial\gamma}
+\sum^{3}_{k=1}\frac{\overline{L}^{2}_{k}}{4sin^{2}(\gamma-2\pi
k/3)} . \label{SO5-Casimir}
\end{equation}

Then the Bohr Hamiltonian takes the well-known form \cite{BM}:
\begin{equation}
H = - \frac{\hbar^{2}}{2B}\nabla^{2} + V(\beta,\gamma),
\label{HBohr}
\end{equation}
where $V(\beta,\gamma)$ is the potential energy. An exact solution
of the Schrodinger equation for the Bohr Hamiltonian (\ref{HBohr})
exist only for few potentials. It is well known that there are three
standard submodels of the BM collective model which are analytically
solvable: one for spherical nuclei, and two for deformed nuclei.

\subsection{Harmonic spherical vibrator}

The Hamiltonian for the spherical vibrator submodel of the BM model
\begin{equation}
H_{HV} = - \frac{\hbar^{2}}{2B}\nabla^{2} + \frac{1}{2}C\beta^{2},
\label{BM-HVH}
\end{equation}
can be expressed, in terms of the creation and annihilation
operators of quadrupole phonons as that of five-dimensional harmonic
oscillator
\begin{equation}
H_{HV} = \sum_{\mu}\bigg(d^{\dag}_{\mu}d_{\mu} +\frac{5}{2}\bigg) =
\bigg(\widehat{N}+ \frac{5}{2}\widehat{I}\bigg), \label{5D-HOH}
\end{equation}
where $\widehat{I}$ is the identity operator. The eigenvectors of
this Hamiltonian define the wave functions, while its eigenvalues
are just the energies and give an equidistant harmonic spectrum with
a characteristic two-phonon multiplet of degenerate $0^{+}$, $2^{+}$
and $4^{+}$ states.

\subsection{Wilets-Jean model}

A major simplification in solving the Schrodinger equation for the
Bohr Hamiltonian (\ref{HBohr}) arises when the potential is
$\gamma$-independent, i.e. $V=V(\beta)$. The WJ model \cite{WJ} is
thus invariant under all $SO(5)$ transformations. The WJ Hamiltonian
is a $SO(5)$ invariant and its eigenvectors occur in multiplets that
span irreducible representations of the $SO(5)$ group. The energies
are labeled by the $SO(5)$ quantum number $\tau$. Recall that the
generic $SO(5)$ irreps are determined by two quantum numbers
$(\tau_{1},\tau_{2})$, but in the case of BM or IBM, they take
one-rowed Young patterns $(\tau_{1},\tau_{2})=(\tau,0)$.

The wave functions of the BM model can be expressed as products of
radial $\beta$ functions and orbital $SO(5)$ wave functions ($SO(5)$
spherical harmonics) \cite{RW}:
\begin{equation}
\Psi_{n\tau\alpha LM}(\beta,\gamma,\Omega) =
R_{n}(\beta)Y_{\tau\alpha LM}(\gamma,\Omega), \label{WJwf}
\end{equation}
where $\tau = 0, 1, 2, \ldots$ is an $SO(5)$ angular momentum
quantum number, which is often referred to as $SO(5)$ seniority. The
explicit expressions for the $SO(5)$ spherical harmonics with $L
\leq 6$ have been obtained by Bes \cite{Bes59} by solving the
coupled system of differential equations that they obey.

The energies and $\beta$ wave functions of the WJ model are
solutions of the eigenvalue equation
\begin{equation}
\bigg[-
\frac{\hbar^{2}}{2B}\bigg(\nabla^{2}-\frac{\tau(\tau+3)}{\beta^{2}}\bigg)
+ V(\beta)\bigg]R_{n}(\beta) = E_{n\tau}R_{n}(\beta). \notag
\end{equation}
A $\beta$-rigid WJ model assumes, in addition, that $\beta$
coordinate is frozen at some non-zero value $\beta_{0}$. Then, the
$\beta$ degree of freedom is suppressed and the Hamiltonian
(\ref{HBohr}) reduces to
\begin{equation}
H_{WJ} = \frac{\hbar^{2}}{2B\beta^{2}_{0}} \Lambda^{2}
\label{WJ-Ham}
\end{equation}
whose eigenvalues are the energies which are given by
\begin{equation}
E_{\tau} = \frac{\hbar^{2}}{2B\beta^{2}_{0}} \tau(\tau+3).
\label{WJen}
\end{equation}
The spectrum for the WJ model is given, for example, in Fig. 6.20 of
Ref.\cite{Casten}. For $\tau > 1$, each value of $\tau$ corresponds
to more than one level and the $\tau$ values $2, 3, 4, \ldots$
include a low-lying set of levels analogous to the
$\gamma$-vibrational band and to the anomalous levels of the
Davydov-Filippov model \cite{DF} for large $\gamma$. The yrast
levels having $L = 2\tau$, according to Eq.(\ref{WJen}), produce a
characteristic ratio $E_{4_{1}}/E_{2_{1}} = 2.50$ of the WJ
$\gamma$-unstable model.

\subsection{Rigid rotor model}

Further simplifications result when both the $\beta$ and $\gamma$
coordinates are frozen. The only remaining degrees of freedom are
then rotations in the three-dimensional space. The $SO(5)$ Casimir
operator (\ref{SO5-Casimir}), with $\beta$ and $\gamma$ taking fixed
values $\beta_{0}$ and $\gamma_{0}$, reduces to
\begin{equation}
\Lambda^{2} =
\sum^{3}_{k=1}\frac{\overline{L}^{2}_{k}}{4sin^{2}(\gamma_{0}-2\pi
k/3)}. \notag
\end{equation}
The BM collective Hamiltonian in this limit correspondingly reduces
to that of a rotor
\begin{equation}
H_{rot} =
\sum^{3}_{k=1}\frac{\hbar^{2}\overline{L}^{2}_{k}}{2J_{k}},
\label{Hrot}
\end{equation}
with the irrotational-flow moments of inertia $J_{k} =
4B\beta^{2}_{0}$ $sin^{2}(\gamma_{0}-2\pi k/3)$. It is known that
the experimental moments of inertia are much larger than these
irrotational flow values. That is why, in numerical applications,
the moments of inertia are treated as free parameters that are
fitted to the experimental data.

\section{Formulation of the BM submodels in algebraic terms. Relation with the IBM}

It is known that almost all models of nuclear structure can be
expressed in algebraic terms of some spectrum generating algebras
and dynamical groups (see, e.g., \cite{Rowe96}). As often happens,
the BM model also has more than one dynamical group. The
Heisenberg-Weyl group $HW(5)=\{\alpha_{\mu},\pi^{\nu},I\}$, which is
the simplest dynamical group spanned by the quadrupole collective
variables and their conjugate momentum operators in
$\mathbb{R}^{5}$, is too small to contain useful subgroup chains
with which to classify the basis states. However, it provides the
basic building blocks from which numerous dynamical groups and
spectrum generating algebras can be constructed. Among them, we
mention the following two dynamical groups: $[HW(5)]U(5)$ and
$SU(1,1) \otimes SO(5)$. The latter turns out to be very efficient
one on which the algebraic version of the BM collective model is
based \cite{Rowe-ACM1,Rowe-ACM2}.

The three BM submodels are associated with dynamical group chains
corresponding to different paths through the set of groups
\cite{Rowe-ACM1,RW}:
\begin{align}
&[HW(5)]U(5) \supset [R^{5}]SO(5) \supset [R^{5}]SO(3) \notag\\
&\qquad\quad \cup \qquad\qquad \ \cup \qquad\qquad \quad \cup \notag\\
&\qquad \ \ U(5) \quad \supset \ SO(5) \quad \ \ \supset \ SO(3),
\label{Hrot}
\end{align}
starting with $[HW(5)]U(5)$ and ending with SO(3), where $R^{5}$ is
the group with Abelian Lie algebra spanned by the quadrupole moments
only, i.e. $R^{5} \equiv
\{\alpha_{\mu}=\frac{1}{\sqrt{2}}(d^{\dag}_{\mu}+d_{\mu})\}$.

\subsection{Harmonic spherical vibrator}

The dynamical subgroup chain for the harmonic vibrator BM submodel
is defined by \cite{Rowe-ACM1}:
\begin{equation}
[HW(5)]U(5) \supset U(5) \supset SO(5) \supset SO(3), \label{HV-DS}
\end{equation}
where $[HW(5)]U(5)$ is the semi-direct product group of
Heisenberg-Weyl group $HW(5)=\{\alpha_{\mu},\pi^{\nu},I\}$ and $U(5)
=\{\alpha_{\mu}\pi^{\nu}+\pi^{\nu}\alpha_{\mu}, \
L_{k}=\sqrt{10}\frac{i}{\hbar}[\alpha \times \pi]_{1k},k=0,\pm1, \
O_{\nu}=\sqrt{10}\frac{i}{\hbar}[\alpha \times
\pi]_{3\nu},\nu=0,\pm1,\pm2,\pm3\}$. In terms of quadrupole phonon
operators, one obtains the alternative realization $[HW(5)]U(5) =
\{d^{\dag}_{\mu}, d_{\nu}, I,$ $d^{\dag}_{\mu}d_{\nu}\}$.

It was shown \cite{RoweThiamova05} that in the spherical vibrator
$U(5)$ limit, the IBM dynamical symmmetry chain \cite{IBM}
\begin{equation}
U(6) \supset U(5) \supset SO(5) \supset SO(3) \label{IBM-U5}
\end{equation}
contracts in the $N\rightarrow \infty$ limit to the BM dynamical
symmetry chain (\ref{HV-DS}), which is actually based on the
Holstein-Primakoff realization of the $U(6)$:
\begin{align}
& s^{\dag}s \rightarrow N, \notag\\
& d^{\dag}_{\mu}s \rightarrow \sqrt{N}d^{\dag}_{\mu}, \notag\\
& s^{\dag}d_{\mu} \rightarrow \sqrt{N}d_{\mu}, \notag\\
& d^{\dag}_{\mu}d_{\nu} \rightarrow d^{\dag}_{\mu}d_{\nu}
\label{HP-U6}
\end{align}
This contraction of the IBM to BM spectrum generating algebra is
valid only for low-energy IBM states close to $U(5)$ limit.
According to this contraction/compactification relation, any
development in the $U(5)$ limit of one model apply equally to the
other.

\subsection{Wilets-Jean model}

The dynamical subgroup chain of the $\gamma$-unstable $\beta$-rigid
WJ model is \cite{Rowe-ACM1}:
\begin{equation}
[HW(5)]U(5) \supset [R^{5}]SO(5) \supset SO(5) \supset SO(3),
\label{WJ-DS}
\end{equation}
where the semi-direct product group $[R^{5}]SO(5)$ consists of an
Abelian ideal $R^{5} \equiv \{\alpha_{\mu};
[\alpha_{\mu},\alpha_{\nu}] = 0 \}$ and the generators of the
$SO(5)$ group. Similarly to the $U(5)$ limit, there is a close
correspondence of the physics of the IBM in its $O(6) \supset SO(5)
\supset SO(3)$ limit with the BM model in its $\beta$-rigid
$\gamma$-unstable WJ limit, as was first shown by J. Meyer-ter-Vehn
\cite{Meyer-ter-Vehn79}. This correspondence is precise in the limit
in which the IBM dynamical symmetry subgroup chain \cite{IBM}
\begin{equation}
U(6) \supset O(6) \supset SO(5) \supset SO(3) \label{IBM-O6}
\end{equation}
contracts in the $N\rightarrow \infty$ limit to the chain
(\ref{WJ-DS}) of the WJ model, based on the $O(6) \rightarrow
[R^{5}]SO(5)$ contraction \cite{RoweThiamova05,Elliott86b}:
\begin{align}
&Q_{\mu}=d^{\dag}_{\mu}s + s^{\dag}d_{\mu} \rightarrow
\sqrt{\upsilon(\upsilon+4)} (d^{\dag}_{\mu}+d_{\mu}), \notag\\
&(d^{\dag} \otimes d_{\nu})_{L} \rightarrow (d^{\dag} \otimes
d_{\nu})_{L}, \qquad L=1,3 \label{O6-WJ}
\end{align}
where $Q_{\mu}=d^{\dag}_{\mu}s + s^{\dag}d_{\mu}$ is the $O(6)$
quadrupole operator and $\upsilon$ is an $O(6)$ irrep. The
irreducible representations of $[R^{5}]SO(5)$ are characterized by
the rigid values of $\beta =\beta_{0}$. Hence, there is a problem
with the delta-function nature of the $\beta$ wave functions, which
in turn don't have a convergent expansion in terms of the harmonic
oscillator $U(5)$ states in the IBM. This problem, as will be
shortly considered further, is circumvented in the ACM
\cite{Rowe-ACM1,Rowe-ACM2} in which the $\beta$-rigid wave functions
of the WJ model $[R^{5}]SO(5)$ algebra are replaced by the
$\beta$-soft wave functions of the dynamical algebra $SU(1,1)
\otimes SO(5)$.

\subsection{Rigid rotor model}

The dynamical subgroup chain of the $\beta$-rigid and $\gamma$-rigid
rotor model is \cite{Rowe-ACM1}:
\begin{equation}
[HW(5)]U(5) \supset [R^{5}]SO(5) \supset [R^{5}]SO(3) \supset SO(3),
\label{RR-DS}
\end{equation}
where $ROT(3) \equiv [R^{5}]SO(3)$ is the rigid-rotor model group of
Ui \cite{rot3}. The irreducible representations of the $ROT(3)$
group are characterized by both $\beta$-rigid and $\gamma$-rigid
values. Thus, in the BM model $\beta$-rigid and $\gamma$-rigid
subgroup chain (\ref{RR-DS}) is a submodel of the $\beta$-rigid but
$\gamma$-unstable subgroup chain (\ref{WJ-DS}) since the
$[R^{5}]SO(3)$ is a subgroup of $[R^{5}]SO(5)$. Hence, in the BM
rigid-rotor submodel again there is a problem with the wave
functions which are delta functions in both $\beta$ and $\gamma$.
Further, the IBM dynamical symmetry limit chain \cite{IBM}
\begin{equation}
U(6) \supset SU(3) \supset SO(3) \label{IBM-SU3}
\end{equation}
is not a submodel of the chain (\ref{IBM-O6}) since the $SU(3)$ is
not a subgroup of $O(6)$. Hence, the rotor-like states in the
$SU(3)$ limit of the IBM are not related to those of its $O(6)$
limit in a way that parallels the relationship between the rigid
rotor and $\beta$-rigid but $\gamma$-unstable states in the BM
model. It is the purpose of the present paper to demonstrate that
there are microscopic shell-model counterparts of the $\beta$-rigid
or $\beta$-soft but $\gamma$-unstable WJ and the $\beta$-rigid and
$\gamma$-rigid rotor limits of the BM model in the configuration
space $\mathbb{R}^{6}$.

\subsection{Algebraic collective model}

Although the last two BM limiting cases just considered are
characterized by dynamical subgroup chains, they are not
particularly useful for the construction of basis states in which to
diagonalize more general collective Hamiltonians, as this is done in
the case of the five-dimensional oscillator. This is because the
wave functions which diagonalize the $[R^{5}]SO(5)$ and
$[R^{5}]SO(3)$ subgroups are not square-integrable. They contain
factors which are delta functions in $\beta$ and $\gamma$. This
limitation expresses the fact that rigidly-defined intrinsic
quadrupole moments are unphysical and incompatible with the quantum
mechanics. The resolution of this problem in the ACM is obtained by
relaxing the $\beta$-rigidity of WJ model by replacing its dynamical
group $[R^{5}]SO(5)$ with $SU(1,1) \otimes SO(5)$, which results in
a more physical collective model. Thus, in the ACM the following
dynamical symmetry chain is used to define a continuous set of basis
states for the BM model \cite{Rowe-ACM1,Rowe-ACM2}:
\begin{align}
&SU(1,1) \otimes SO(5) \supset U(1) \otimes SO(3) \supset SO(2),
\label{ACM-DS} \\
&\qquad \lambda_{\upsilon} \qquad\quad \ \ \upsilon \quad \ \alpha
\quad n \qquad\quad L \qquad\quad M
\end{align}
where $SU(1,1)$ is a dynamical group for radial $\beta$ wave
functions and $SO(5)$ group determines the angular part ($SO(5)$
spherical harmonics) that is characterized by the seniority quantum
number $\upsilon$. An important characteristic of the ACM is that it
enables $\beta$-rigid and $\gamma$-rigid limits to be approached in
a continuous way with increasingly narrow but nevertheless
square-integrable $\beta$ and $\gamma$ wave functions. It turns out
that the convergence for deformed nuclei in the ACM is much faster
than in the states of conventional five-dimensional oscillator
\cite{Rowe-ACM2}. The WJ and rigid-rotor submodels of the BM model
are then seen as special cases of the more physical ACM.

\section{Embedding of the Bohr-Mottelson model in the microscopic shell-model theory}

As we mentioned in the Introduction, the problem of incorporating
the BM collective model into the microscopic shell-model theory of
the nucleus has been realized long time ago. As is known, the
solution \cite{stretched,Rowe96} of this problem was given through
the algebraic approach. Within the latter, the embedding problem
becomes straightforward once it is recognized that both the
collective model of interest and the shell model are algebraic
models with dynamical groups. Then, a given phenomenological
collective model becomes a submodel of the shell model if its
dynamical group is expressed as a subgroup of a dynamical group of
the shell model.

\subsection{Embedding in the one-component microscopic shell-model theory}

It is known that in its standard formulation, the BM collective
model can not be naturally related to the microscopic theory of
nucleus because the vectors in the BM model cannot be identified
with the wave functions in the many-particle Hilbert space of $A$
nucleon antisymmetry states. This is so, because its dynamical group
$[HW(5)]U(5)$ is not the most appropriate. In this respect, we
recall briefly the embedding of the BM model into the one-component
microscopic shell-model theory of nuclear collective motions
obtained many years ago (see, e.g., \cite{stretched,Rowe96}).

The first step in the progression to a microscopic collective model
\cite{stretched,Rowe96,RW} is the replacement of the surface shape
coordinates $\{\alpha_{\nu}\}$, which don't have a microscopic
interpretation and also cause the non-square integrability of rigid
rotor wave functions, by microscopic Cartesian components of the
mass quadrupole tensor
\begin{equation}
Q_{ij} = \sum_{s=1}^{A}x_{is}x_{js}, \quad i,j = 1, 2, 3; s = 1,
\ldots, A . \label{QM}
\end{equation}
It immediately follows then that the time derivatives of the
quadrupole moments and corresponding momentum observables are given
by
\begin{align}
&\dot{Q}_{ij} = \frac{dQ_{ij}}{dt} =
\sum_{s=1}^{A}(\dot{x}_{is}x_{js} +
x_{is}\dot{x}_{js}),\\
&P_{ij} = M\dot{Q}_{ij} = \sum_{s=1}^{A}(p_{is}x_{js} +
x_{is}p_{js}) \neq -i\hbar\frac{\partial}{\partial Q_{ij}},
\label{QdotP}
\end{align}
where $M$ is the nucleon mass. These moments and momenta are
quantized by replacing the $x_{is}$ and $p_{is}$ coordinates by
operators $\hat{x}_{is}$ and $\hat{p}_{is}$ with commutation
relations $[\hat{x}_{is}, \hat{p}_{jt}] = i\hbar
\delta_{ij}\delta_{st}$ to obtain the quantum observables
\begin{equation}
\hat{Q}_{ij} = \sum_{s=1}^{A}\hat{x}_{is}\hat{x}_{js}, \quad
\hat{P}_{ij} = \sum_{s=1}^{A}(\hat{p}_{is}\hat{x}_{js} +
\hat{x}_{is}\hat{p}_{js}). \label{QPhat}
\end{equation}
From (\ref{QdotP}) it becomes clear that the quantization of the BM
model given by the standard Heisenberg-Weyl commutation relations
\begin{equation}
[\alpha_{\mu},\pi^{\nu}] = i\hbar \delta^{\nu}_{\mu},
\label{HW5comm}
\end{equation}
where $\pi^{\nu} = -i\hbar\frac{\partial}{\partial \alpha_{\nu}}$,
was not correct. The new commutation relations emerge
\begin{equation}
[\hat{Q}_{ij},\hat{P}_{kl}] = i\hbar\big(\delta_{il}\hat{Q}_{jk} +
\delta_{ik}\hat{Q}_{jl} + \delta_{jl}\hat{Q}_{ik} +
\delta_{jk}\hat{Q}_{il}\big), \label{QPcomm}
\end{equation}
which together with the antisymmetric angular momentum operators
$\hbar \hat{L}_{k} = \hbar \epsilon_{kij}\hat{L}_{ij} =
\sum_{s=1}^{A} (\hat{x}_{si}\hat{p}_{sj}$
$-\hat{x}_{sj}\hat{p}_{si})$ span the Lie algebra of general
collective motion group in three dimensions, i.e. $GCM(3) =
\{\hat{L}_{ij}, \hat{Q}_{ij}, \hat{P}_{ij}\}$. For simplicity,
further the hats in the notations of generators will be suppressed.
The $GCM(3)$ model is slightly extended version of the original
$CM(3)$ model of Weaver, Biedenharn and Cusson \cite{cm3,NA4}, which
in addition includes the monopole degrees of freedom. The $CM(3)$
model was obtained by extending the motion group (i.e. the group of
transformations of the configuration space) from $SO(3)$ to
$SL(3,R)$ which includes into considerations beyond the rotational
also the vibrational degrees of freedom of shape change. What is
remarkable is that the new spectrum generating algebra of the
$GCM(3)$ model has irreps with different intrinsic angular momenta
(vorticities). It was shown by G. Rosensteel \cite{Rosensteel88}
that the invariant operator of the $GCM(3)$ model is represented by
the square of conserved Kelvin circulation vector, i.e. $V^{2}$, and
its eigenvalues $V(V+1)$ correspond to the quantized vorticity. The
$V = 0$ representation has states that are in one-to-one
correspondence with those of the BM model. In this way the $GCM(3)$
model extends the irrotational-flows of the BM model to include an
$SO(3)$ intrinsic gauge (vorticity) degrees of freedom that are
important for the appearance of low-lying collective states in
nuclear spectra.

In addition of being a microscopic version of the BM model augmented
by the intrinsic vortex degrees of freedom, the $GCM(3)$ model has
the desirable characteristic of containing all physical observables,
i.e., quadrupole moments, angular momenta, vortex spin, and
infinitesimal generators of deformation, that appear in the
expression of the collective component $T_{coll}$ \cite{Rowe96} of
the many-nucleon kinetic energy. As shown by Rosensteel, the
$GCM(3)$ is also related to the Riemann model of rotating ellipsoids
\cite{Rosensteel88} with linear combinations of rigid and
irrotational flows and has a mathematical structure in terms of
Yang-Mills theory \cite{Rosensteel17}. The problem with the $GCM(3)$
model is that it is not compatible with the shell model. The irreps
of $GCM(3)$ have no simple shell-model expression, except for the
trivial case of vortex-free irreps. Additionally, the kinetic energy
of the $GCM(3)$ model has an exceedingly complicated expression and
the model is difficult to use in the calculations of nuclear
properties with a many-nucleon Hamiltonian. But the more serious
concern is that the full many-particle kinetic energy does not
conserve the vortex spin, i.e., it strongly mixes different $GCM(3)$
irreps.

A resolution of the problem with the $GCM(3)$ model was obtained by
simply extending it to the one-component symplectic $Sp(6,R)$ model
\cite{RR1} which includes the full many-particle kinetic energy $T =
\sum_{s=1}^{A}\hat{\textbf{p}}^{2}_{s}/2M$. The Lie algebra that
emerges then contains all symmetric bilinear combinations of the
many-nucleon position $\{x_{is}\}$ and momentum $\{p_{is}\}$
coordinates. The SGA of the $Sp(6,R)$ model thus becomes
\cite{stretched,Rowe96,RW}:
\begin{align}
&Q_{ij} = \sum_{s=1}^{A}x_{is}x_{js}, \quad P_{ij} =
\sum_{s=1}^{A}(p_{is}x_{js} + x_{is}p_{js}), \\
&\hbar L_{ij} = \sum_{s=1}^{A}(x_{si}p_{sj}-x_{sj}p_{si}), \quad
K_{ij} = \sum_{s=1}^{A}p_{is}p_{js}. \label{Sp6R-gen}
\end{align}
$Sp(6,R)$ is also the smallest Lie algebra that contains the nuclear
quadrupole moments and the many-nucleon kinetic energy, both of
which are essential components of a complete microscopic model of
nuclear collective states. In addition, the $Sp(6,R)$ SGA contains
the infinitesimal generators of both rigid- and irrotational-flow
rotations. It contains also the $U(3)$ Lie algebra of the Elliott
model as a subalgebra and has the valuable property that it defines
a coupling scheme in a $U(3) \supset SU(3)$ basis for the
many-nucleon Hilbert space in a straightforward way. The $Sp(6,R)$
Lie algebra, like that of $U(3)$, can be augmented to include the
$U(4)$ supermultiplet spin-isospin algebra with which it commutes.
The one-component $Sp(6,R)$ model then defines a complete coupling
scheme for the many-nucleon shell-model Hilbert space in a spherical
harmonic-oscillator basis. An important property of the $Sp(6,R)$
model is that it also appears as a bridge between the shell model
and the collective model. Thus, it is a microscopic unified model in
every respect. Moreover, as we will see in the next subsection, the
$Sp(6,R)$ SGA can be embedded in the $Sp(12,R)$ dynamical algebra,
i.e. $Sp(6,R) \subset Sp(12,R)$, which allows for the more complete
description of the complex proton-neutron dynamics, as well as the
separate treatment of the proton and neutron subdynamics. As a
result, a new fully microscopic model of collective excitations in
the two-component many-particle nuclear systems arises, namely the
proton-neutron symplectic model \cite{cdf,smpnsm}. The latter will
allow us to give the BM collective model a microscopic foundation,
admitting a more natural interpretation of the underlying BM
quadrupole-monopole collective dynamics that is missing in the
(one-component) $Sp(6,R)$ symplectic model.

\subsection{Embedding in the two-component proton-neutron microscopic shell-model theory}

In the present subsection we will consider the embedding of the
generalized BM collective model into the microscopic shell-model
theory of atomic nucleus in the framework of the proton-neutron
symplectic model, which provides a more natural interpretation of
the underlying BM quadrupole-monopole collective dynamics than the
embedding given in the preceding subsection.

The PNSM with $Sp(12,R)$ SGA was formulated by considering the
symplectic geometry and possible collective flows in the
two-component many-particle nuclear system \cite{cdf}, generalizing
in this way the $Sp(6,R)$ model. The PNSM collective observables are
given by the following $O(A-1)$-invariant one-body operators
\cite{cdf}:
\begin{eqnarray}
&&Q_{ij}(\alpha,\beta)=\sum_{s=1}^{m}x_{is}(\alpha)x_{js}(\beta),
\label{Sp12a} \\
&&S_{ij}(\alpha,\beta)=
\sum_{s=1}^{m}\biggl(x_{is}(\alpha)p_{js}(\beta)+p_{is}(\alpha)x_{js}(\beta)\biggr),
\label{Sp12b} \\
&&L_{ij}(\alpha,\beta)=\sum_{s=1}^{m}\biggl(x_{is}(\alpha)p_{js}(\beta)-x_{js}(\beta)p_{is}(\alpha)\biggr),
\label{Sp12c} \\
&&T_{ij}(\alpha,\beta)=\sum_{s=1}^{m}p_{is}(\alpha)p_{js}(\beta),
\label{Sp12d}
\end{eqnarray}
where  $i,j = 1,2,3$; $\alpha,\beta = p,n$ and $s = 1,\ldots,m=A-1$.
In Eqs.(\ref{Sp12a})$-$(\ref{Sp12d}), $x_{is}(\alpha)$ and
$p_{is}(\alpha)$ denote the coordinates and corresponding momenta of
the translationally-invariant Jacobi vectors of the
$m$-quasiparticle two-component nuclear system and $A$ is the number
of protons and neutrons. By considering the $m$ Jacobi
quasiparticles instead of $A$ protons and neutrons, the problem of
center-of-mass motion is avoided from the very beginning. Obviously,
by summing over $\alpha$ in Eqs.(\ref{Sp12a})$-$(\ref{Sp12d}) we
obtain the one-component $Sp(6,R)$ symplectic model as a submodel.

The symplectic generators of the PNSM can be written in an
alternative form in terms of all bilinear combinations of the
raising and lowering operators of harmonic oscillator quanta
\begin{align}
&b^{\dagger}_{i\alpha,s}=
\sqrt{\frac{m_{\alpha}\omega}{2\hbar}}\Big(x_{is}(\alpha)
-\frac{i}{m_{\alpha}\omega}p_{is}(\alpha)\Big), \notag\\
&b_{i\alpha,s}=\sqrt{\frac{m_{\alpha}\omega}{2\hbar}}\Big(x_{is}(\alpha)
+\frac{i}{m_{\alpha}\omega}p_{is}(\alpha)\Big). \label{bos}
\end{align}
that are $O(m)$ invariant \cite{smpnsm}:
\begin{align}
&F_{ij}(\alpha,\beta)=\sum_{s=1}^{m}b^{\dagger}_{i\alpha,s}b^{\dagger}_{j\beta,s},
\label{Fs} \\
&G_{ij}(\alpha,\beta)=\sum_{s=1}^{m}b_{i\alpha,s}b_{j\beta,s},
\label{Gs} \\
&A_{ij}(\alpha,\beta)=\frac{1}{2}\sum_{s=1}^{m}
(b^{\dagger}_{i\alpha,s}b_{j\beta,s}+b_{j\beta,s}b^{\dagger}_{i\alpha,s}).
\label{As}
\end{align}
The operators $A_{ij}(\alpha,\beta)$ are the generators of the
maximal compact subgroup $U(6) \subset Sp(12,R)$. We introduce also
the operators \cite{wsp12r}:
\begin{equation}
B^{\dag}_{i}(\alpha)= \sum_{s}b^{\dagger}_{i\alpha,s} \label{Bdag}
\end{equation}
and $B_{i}(\alpha)=\big(B^{\dag}_{i}(\alpha)\big)^{\dag}$, which
together with the identity operator close the six-dimensional
Heisenberg-Weyl algebra $HW(6) = \{B^{\dag}_{i}(\alpha),
B_{i}(\alpha),I\}$.

The set of operators $\{L_{ij}(\alpha,\beta),S_{ij}(\alpha,\beta)\}$
form the Lie algebra of a more general motion group $GL(6,R) \subset
Sp(12,R)$, which allows for the separate treatment of the collective
dynamics of proton and neutron subsystems as well as the combined
proton-neutron collective excitations. The configuration space of
the PNSM is isomorphic to the coset space $GL(6,R)/SO(6)$ and is
spanned by the commuting quadrupole momentum observables, i.e.
$\mathbb{R}^{21} = \{Q_{ij}(\alpha,\beta)\}$. Moreover,
Eq.(\ref{Sp12a}) can be considered as a map from the microscopic
many-particle configuration space $\mathbb{R}^{6m}$ to the
collective configuration space \cite{cdf}:
\begin{equation}
Q: \mathbb{R}^{6m} \rightarrow \mathbb{Q}; \quad x \rightarrow Q(x)
= \widetilde{x}x, \label{Qmap}
\end{equation}
where $\widetilde{x}$ denotes the transpose of the matrix $x \in
\mathbb{R}^{6m}$. It follows that every path $x(t)$ in
$\mathbb{R}^{6m}$ has an image $Q(x(t))$ in $\mathbb{Q}$. Thus, the
collective motions in $\mathbb{R}^{6m}$ map to collective motions in
$\mathbb{Q}$. In this way the components of the quadrupole moment
(\ref{Sp12a}) define the microscopic collective configuration space
$\mathbb{R}^{21}$ of the PNSM.

Now, let us consider instead of the mapping $\{\alpha_{\nu}\}
\rightarrow \{Q_{ij}\}$ the following one: $\{\alpha_{\nu}\}
\rightarrow \{Q_{ij}(p,n)\}$. Then, in contrast to the $Sp(6,R)$
case, we obtain a six-dimensional microscopic configuration subspace
$\mathbb{R}^{6} \subset \mathbb{R}^{21}$, spanned by the six
commuting components $\{Q_{ij}(p,n)=Q_{ji}(n,p)\}$, in which the
group of six-dimensional rigid rotations $SO(6)$ acts. As will be
demonstrated further, this configuration space $\mathbb{R}^{6} =
\{Q_{ij}(p,n)\}$ is isomorphic to the configuration collective space
of the generalized quadrupole-monopole Bohr-Mottelson dynamics.
Further, the same kind of considerations are valid as those for the
case of one-component nuclear systems given in the previous
subsection.

The $SO(6)$ group, spanned by the components of the six-dimensional
angular-momentum operators $\{L_{ij}(\alpha,\beta)\}$, can be
expressed more conveniently in terms of the $U(6)$ generators
(\ref{As}) as \cite{sp2rxso6}:
\begin{equation}
\Lambda^{LM}(\alpha,\beta)=A^{LM}(\alpha,\beta)
-(-1)^{L}A^{LM}(\beta,\alpha).  \label{O6gen}
\end{equation}
It then allows us to consider the following reduction chain of the
PNSM:
\begin{equation}
Sp(12,R) \supset U(6) \supset SO(6) \supset G \supset SO(3). \notag
\end{equation}
The reduction of $SO(6)$ to $SO(3)$ can be carried out in different
ways, but for the present purposes we choose $G = SU(3) \otimes
SO(2)$ \cite{Dragt65,Chacon82,Chacon84,LeBlanc86}. Thus, we obtain
the subgroup chain
\begin{equation}
Sp(12,R) \supset U(6) \supset SO(6) \supset SU_{pn}(3) \otimes SO(2)
\supset SO(3), \label{O6-DS}
\end{equation}
which defines a microscopic shell-model analogue of the BM
$\gamma$-unstable reduction chain. We point out that the PNSM
$Sp(12,R)$ Lie algebra, realized in terms of many-particle position
and momentum Jacobi coordinates, is a subalgebra of the shell-model
algebra of one-body unitary transformations. Thus the problem of
embedding of the generalized BM quadrupole-monopole collective
dynamics of WJ-type into the microscopic shell-model theory of a
nucleus is solved from the beginning.

The second-order invariant for the $SO(6)$\ group is
\begin{equation}
\Lambda^{2} = \sum_{\alpha,\beta,L,M} (-1)^{M}
\Lambda^{LM}(\alpha,\beta)\Lambda^{L,-M}(\beta,\alpha). \label{C2O6}
\end{equation}
The generators of the $SU_{pn}(3)$ group are defined by
\cite{sp2rxso6}
\begin{align}
&\widetilde{q}^{2M}= \sqrt{3} i[A^{2M}(p,n)-A^{2M}(n,p)],
\label{X-SUpn3}
\\
&Y^{1M}=\sqrt{2}[A^{1M}(p,p)+A^{1M}(n,n)], \label{SUpn3gen}
\end{align}
whereas the single infinitesimal operator of $SO(2)$ is proportional
to the $SO(3)$ scalar operator $\Lambda^{0}(\alpha,\beta)$
\cite{sp2rxso6}:
\begin{equation}
M_{\alpha,\beta}=\Lambda^{0}(\alpha,\beta) =
i[A^{0}(\alpha,\beta)-A^{0}(\beta,\alpha)]. \label{O2gen}
\end{equation}
Obviously, by construction the generator of $SO(2)$ commute with the
generators of $SU_{pn}(3)$ which are rank-2 and rank-1 tensors. The
reduction of the $SO(6)$ to $SO(3)$ is therefore carried out by two
mutually complementary groups $SU_{pn}(3)$ and $SO(2)$ \cite{MQ70},
i.e. we have a direct-product group $SU_{pn}(3) \otimes SO(2)$. Note
also that in the present case the quadrupole momentum operators
$\widetilde{q}$ (\ref{X-SUpn3}) are of proton-neutron nature. The
second-order Casimir operators of the two groups $SU_{pn}(3)$ and
$SO(2)$ are given by \cite{sp2rxso6}:
\begin{align}
&C_{2}[SU_{pn}(3)] =
\sum_{M}(-1)^{M}(\widetilde{q}^{2M}\widetilde{q}^{2,-M}
+Y^{1M}Y^{1,-M}),
\notag \\
&C_{2}[SO(2)] = M^{2} =
\sum_{\alpha,\beta}M_{\alpha\beta}M_{\beta\alpha} . \label{C2SUpn3}
\end{align}
From the above expressions, it follows that the second-order Casimir
operator of $SO(6)$ can be written in an alternative form as:
\begin{equation}
C_{2}[SO(6)]= (\widetilde{q} \cdot \widetilde{q} + Y \cdot Y)
+L^{2}_{p}+L^{2}_{n}. \label{C2O6a}
\end{equation}
and its eigenvalue $\upsilon(\upsilon+4)$ is determined by the
quantum number $\upsilon$ characterizing the $SO(6)$ irreps.

For $SO(6)\subset U(6)$, the symmetric representation $[N]_{6}$ of
$U(6)$ decomposes into fully symmetric $(\upsilon ,0,0)_{6}\equiv
(\upsilon )_{6}$ irreps of $SO(6)$ according to the rule
\cite{Van71,sp2rxso6}:
\begin{equation}
\lbrack N]_{6}=\bigoplus_{\upsilon =N,N-2,...,0(1)}(\upsilon
,0,0)_{6}=\bigoplus_{i=0}^{<\frac{N}{2}>}(N-2i)_{6}.  \label{U6O6}
\end{equation}
Furthermore, the following relation between the quadratic \ Casimir
operators $C_{2}[SU_{pn}(3)]$ of $SU_{pn}(3)$, $M^{2}$ of $SO(2)$
and $C_{2}[SO(6)]$  of $SO(6) $ holds
\cite{Dragt65,Chacon84,sp2rxso6}:
\begin{equation}
C_{2}[SO(6)]=2C_{2}[SU_{pn}(3)]-\frac{1}{3}M^{2}.
\label{O6SU3O2rel}
\end{equation}
As a consequence, the irrep labels $[f_{1},f_{2},0]_{3}$ of
$SU_{pn}(3)$ are determined by $(\upsilon )_{6}$ of $SO(6)$\ and by
the integer label $(\nu )_{2}$ of the associated irrep of $SO(2)$
i.e.
\begin{equation}
(\upsilon )_{6}=\bigoplus [f_{1},f_{2},0]_{3}\otimes (\nu )_{2}.
\label{o6su3qn}
\end{equation}%
Using the relation (\ref{O6SU3O2rel}) of the Casimir operators, for
their respective eigenvalues one obtains:
\begin{equation}
\upsilon (\upsilon
+4)=\frac{4}{3}(f_{1}^{2}+f_{2}^{2}-f_{1}f_{2}+3f_{1})
-\frac{\nu^{2}}{3}. \notag
\end{equation}
Thus (\ref{o6su3qn}) can be rewritten as
\begin{align}
(\upsilon )_{6} &=\bigoplus_{i=0}^{\upsilon }[\upsilon
,i,0]_{3}\otimes (\upsilon -2i)_{2} \notag\\
\notag\\
&=\bigoplus_{\nu=\upsilon ,\upsilon -2,...,0(1)}[\upsilon
,\frac{\upsilon -\nu }{2},0]_{3}\otimes (\nu)_{2}, \notag
\end{align}
or in terms of the Elliott's notation $(\lambda,\mu)$:
\begin{equation}
(\upsilon )_{6}=\bigoplus_{\nu =\upsilon ,\upsilon
-2,...,0(1)}(\lambda=\frac{\upsilon +\nu }{2},\mu=\frac{\upsilon
-\nu }{2})\otimes (\nu )_{2}.  \label{O6SUpn3}
\end{equation}
Finally, the convenience of this reduction can be further enhanced
through the use of the standard rules for the reduction of the
$SU_{pn}(3) \supset SO(3)$ chain in terms of a multiplicity index
$q$ which distinguishes the same $L$ values in the $SU_{pn}(3)$
multiplet $(\lambda,\mu)$ \cite{Elliott58,sp2rxso6}:
\begin{eqnarray}
q &=&\min(\lambda,\mu),\min(\lambda,\mu)-2,...,0~(1)  \notag \\
L &=&\max(\lambda,\mu ),\max(\lambda,\mu)-2,...,0~(1); \ q=0 \qquad
\label{su3o3}
\\
L &=&q,q+1,...,q+\max(\lambda,\mu); \ q\neq 0.  \notag
\end{eqnarray}

Using the above reduction rules we give in Table \ref{SUpn3BS} the
$SU_{pn}(3)$ basis states for first few even $SO(6)$ irreps,
starting from zero, which correspond to the lowest positive-parity
states of spherical doubly closed-shell nuclei.

\begin{table}[h]
\caption{The $SU_{pn}(3)$ basis states in the $SO(6)$ dynamical
symmetry limit (\ref{O6-DS}) of the PNSM, obtained according to the
reduction rules defined by Eqs.(\ref{U6O6}) and (\ref{O6SUpn3}),
respectively.} \label{SUpn3BS}
\smallskip\centering\small\addtolength{\tabcolsep}{-0.5pt}
\begin{tabular}{||c|c|ccccccccc||}
\hline\hline $N$ & $\upsilon $ & $\nu \cdots $ & $\quad 6$ & $4$ &
$2$ & $0$ & $-2$ & $-4$ & $-6\quad $ & $\cdots $ \\ \hline\hline
\multicolumn{1}{||l|}{$\vdots $} & $\vdots $ &
\multicolumn{1}{|l}{$\vdots $} & \multicolumn{1}{l}{ $\ \ \ \ \
\vdots $} & \multicolumn{1}{l}{ $\ \ \ \ \
\vdots $} & \multicolumn{1}{l}{ $\ \ \ \ \ \vdots $} & \multicolumn{1}{l}{ $%
\ \ \ \ \ \vdots $} & \multicolumn{1}{l}{ $\ \ \ \ \ \vdots $} &
\multicolumn{1}{l}{ $\ \ \ \ \ \vdots $} & \multicolumn{1}{l}{ $\ \
\ \ \vdots $} & \multicolumn{1}{l||}{} \\ \hline
\multicolumn{1}{||l|}{$6$} &
\begin{tabular}{l}
$6$ \\
$4$ \\
$2$ \\
$0$%
\end{tabular}
& \multicolumn{1}{|l}{} & \multicolumn{1}{l}{%
\begin{tabular}{l}
$(6,0)$ \\
\\
\\
.%
\end{tabular}%
} & \multicolumn{1}{l}{%
\begin{tabular}{l}
$(5,1)$ \\
$(4,0)$ \\
\\
.%
\end{tabular}%
} & \multicolumn{1}{l}{%
\begin{tabular}{l}
$(4,2)$ \\
$(3,1)$ \\
$(2,0)$ \\
.%
\end{tabular}%
} & \multicolumn{1}{l}{%
\begin{tabular}{l}
$(3,3)$ \\
$(2,2)$ \\
$(1,1)$ \\
$(0,0)$%
\end{tabular}%
} & \multicolumn{1}{l}{%
\begin{tabular}{l}
$(2,4)$ \\
$(1,3)$ \\
$(0,2)$ \\
.%
\end{tabular}%
} & \multicolumn{1}{l}{%
\begin{tabular}{l}
$(1,5)$ \\
$(0,4)$ \\
\\
.%
\end{tabular}%
} & \multicolumn{1}{l}{%
\begin{tabular}{l}
$(0,6)$ \\
\\
\\
.%
\end{tabular}%
} & \multicolumn{1}{l||}{} \\ \hline \multicolumn{1}{||l|}{$4$} &
\begin{tabular}{l}
$4$ \\
$2$ \\
$0$%
\end{tabular}
& \multicolumn{1}{|l}{} & \multicolumn{1}{l}{} & \multicolumn{1}{l}{%
\begin{tabular}{l}
$(4,0)$ \\
\\
.%
\end{tabular}%
} & \multicolumn{1}{l}{%
\begin{tabular}{l}
$(3,1)$ \\
$(2,0)$ \\
.%
\end{tabular}%
} & \multicolumn{1}{l}{%
\begin{tabular}{l}
$(2,2)$ \\
$(1,1)$ \\
$(0,0)$%
\end{tabular}%
} & \multicolumn{1}{l}{%
\begin{tabular}{l}
$(1,3)$ \\
$(0,2)$ \\
.%
\end{tabular}%
} & \multicolumn{1}{l}{%
\begin{tabular}{l}
$(0,4)$ \\
\\
.%
\end{tabular}%
} & \multicolumn{1}{l}{} & \multicolumn{1}{l||}{} \\ \hline
\multicolumn{1}{||l|}{$2$} &
\begin{tabular}{l}
$2$ \\
$0$%
\end{tabular}
& \multicolumn{1}{|l}{} & \multicolumn{1}{l}{} &
\multicolumn{1}{l}{} &
\multicolumn{1}{l}{%
\begin{tabular}{l}
$(2,0)$ \\
.%
\end{tabular}%
} & \multicolumn{1}{l}{$%
\begin{tabular}{l}
$(1,1)$ \\
$(0,0)$%
\end{tabular}%
\ \ $} & \multicolumn{1}{l}{%
\begin{tabular}{l}
$(0,2)$ \\
.%
\end{tabular}%
} & \multicolumn{1}{l}{} & \multicolumn{1}{l}{} & \multicolumn{1}{l||}{} \\
\hline \multicolumn{1}{||l|}{$0$} & $0$ & \multicolumn{1}{|l}{} &
\multicolumn{1}{l}{} & \multicolumn{1}{l}{} & \multicolumn{1}{l}{} &
\multicolumn{1}{l}{ $ (0,0)$} & \multicolumn{1}{l}{} &
\multicolumn{1}{l}{ } & \multicolumn{1}{l}{} &
\multicolumn{1}{l||}{} \\ \hline\hline
\end{tabular}%
\end{table}

\section{Microscopic counterparts of the exactly solvable limits of the BM model}

We are now ready to define the microscopic many-particle
counterparts of the generalized quadrupole-monopole BM submodels,
which closely parallels the relationship between the original
Wilets-Jean and rotor models given in Section II.

\subsection{Microscopic counterpart of the WJ model}

Formally, further we can form the six-dimensional analogue of BM
dynamical group: $[HW(6)]U(6) =\{B^{\dag}_{i}(\alpha),$
$B_{i}(\alpha), I, A_{ij}(\alpha,\beta)\}$ which is a semi-direct
group of the $HW(6)$ and $U(6)$. Then, we can consider the following
reduction chain
\begin{align}
[HW(6)]U(6) &\supset [R^{6}]SO(6) \notag\\
\notag\\
&\supset SO(6) \supset SU(3) \otimes SO(2) \supset SO(3),
\label{6DWJ-DS}
\end{align}
where the semi-direct product group $[R^{6}]SO(6)$ consists of an
Abelian ideal $R^{6} \equiv \{x_{i}(\alpha);
[x_{i}(\alpha),x_{j}(\beta)] = 0 \}$ and the generators of the
$SO(6)$ (\ref{O6gen}), i.e. $[R^{6}]SO(6) \equiv \{x_{i}(\alpha) =
\frac{1}{\sqrt{2}}[B^{\dag}_{i}(\alpha)
+B_{i}(\alpha)],\Lambda^{LM}(\alpha,\beta)\}$ group. This
construction parallels that of the $\beta$-rigid WJ model and the
irreps of $[R^{6}]SO(6)$ are characterized by a fixed value $r_{0}$
of the radial coordinate (hyper-radius)
$r=\sqrt{r^{2}_{p}+r^{2}_{n}}$, where $r^{2}_{\alpha}= \sum_{s}
x^{2}_{s}(\alpha)$ and $\alpha=p,n$. The square of the hyper-radius
$r^{2}$ is invariant under $SO(6)$ transformations.

\subsection{Microscopic counterpart of the rigid-rotor model}

An analogue of the rigid-rotor submodel of the BM model, defined by
the dynamical group chain (\ref{RR-DS}), can be defined within the
framework of the PNSM by considering the following subgroup chain
\begin{align}
[HW(6)]U(6) &\supset [R^{6}]SO(6) \notag\\
\notag\\
&\supset SO(6) \supset [R^{5}]SO(3) \supset SO(3),
\label{6DWJ-RR-DS}
\end{align}
where the dynamical group $[R^{5}]SO(3)$ of the rigid rotor is
easily obtained as a contraction limit of $SU_{pn}(3)$ for large
$SU(3)$ representations \cite{su3rot3}.

\section{Many-particle quantum-mechanical shell-model counterpart of the BM model}
\label{SU11xSO6-SM}

Similarly to the case of $\beta$-rigid BM submodels, it is more
convenient to relax the rigidity constrain and to consider the
following reduction chain
\begin{align}
SU(1,1) \otimes SO(6) \supset U(1) \otimes SU_{pn}(3) \otimes SO(2)
\supset SO(3), \notag
\end{align}
which is naturally contained in the $Sp(12,R)$ dynamical group of
PNSM. To see this, one just needs to take into account that
$Sp(2,R)$ is locally isomorphic to $SU(1,1)$. We can then define the
following dynamical symmetry limit of the PNSM \cite{sp2rxso6}:
\begin{align}
Sp(12,R) &\supset Sp(2,R) \otimes SO(6) \notag\\
&\qquad\qquad\quad \wr \wr \notag\\
&\supset SU(1,1) \otimes SO(6) \notag\\
&\qquad\quad \lambda_{\upsilon} \qquad\quad \upsilon \notag\\
&\supset U(1) \otimes SU_{pn}(3) \otimes SO(2) \supset SO(3),
\label{Sp2RxO6-DS}\\
&\qquad \ n \qquad \ (\lambda,\mu) \qquad\quad \nu \quad \ q \quad \
L \notag
\end{align}
which is well defined in the many-nucleon quantum mechanics and
completely avoids the problem of non-normalizable wave functions of
the WJ-type models in the zero-width limit in which they become
proportional to a delta function. We note that because of the dual
pair relationships, dynamical symmetry chain (\ref{Sp2RxO6-DS}) is
equivalent to that defined by Eq.(\ref{O6-DS}).

The $Sp(2,R) \approx SU(1,1)$ is a dynamical group for radial wave
functions and $SO(6)$ group determines the angular part ($SO(6)$
spherical harmonics) that is characterized by the seniority quantum
number $\upsilon$. The infinitesimal generators of the $SU(1,1)$
group are expressed in terms of the symplectic generators
(\ref{Fs})$-$(\ref{As}) in the form \cite{sp2rxso6}:
\begin{align}
S^{(\lambda_{\upsilon})}_{+} = \frac{1}{2}\sum_{\alpha}
F^{0}(\alpha,\alpha),
\label{Sp}\\
S^{(\lambda_{\upsilon})}_{-} = \frac{1}{2}\sum_{\alpha}
G^{0}(\alpha,\alpha),
\label{S-}\\
S^{(\lambda_{\upsilon})}_{0} = \frac{1}{2}\sum_{\alpha}
A^{0}(\alpha,\alpha), \label{S0}
\end{align}
for any value of $\lambda_{\upsilon}$ ($\lambda_{\upsilon}
> 1$) which, generally, define the so called modified oscillator $SU(1,1)$
irreps \cite{Rowe-Euclidean}. The wave functions can therefore be
expressed as products of radial $r$ functions and orbital $SO(6)$
wave functions \cite{sp2rxso6}:
\begin{equation}
\Psi_{\lambda_{\upsilon}n;\upsilon\nu qLM}(r,\Omega_{5}) =
R^{\lambda_{\upsilon}}_{n}(r)Y^{\upsilon}_{\nu qLM}(\Omega_{5}),
\label{SU11xSO6wf}
\end{equation}
where $Y^{\upsilon}_{\nu qLM}(\Omega_{5})$ are the $SO(6)$ Dragt's
spherical harmonics \cite{Dragt65,Chacon84}.

The configuration space of the PNSM model in the present case is the
six-dimensional Euclidean space $\mathbb{R}^{6}$. The volume element
in spherical coordinates is given by
\begin{equation}
dV = r^{5}dr d\Omega_{5}, \notag
\end{equation}
where $d\Omega_{5}$ is the volume element of the five-sphere. The
Laplacian operator is
\begin{equation}
\nabla^{2} = \frac{1}{r^{5}}\frac{\partial}{\partial
r}r^{5}\frac{\partial}{\partial r} -\frac{\Lambda^{2}}{r^{2}},
\label{Laplacian}
\end{equation}
where the $SO(6)$ Casimir operator $\Lambda^{2}$ was given by
Eq.(\ref{C2O6}). The energies and radial wave functions can then be
found as solutions of the eigenvalue equation
\begin{equation}
\bigg[-
\frac{\hbar^{2}}{2m}\bigg(\nabla^{2}-\frac{\upsilon(\upsilon+4)}{r^{2}}\bigg)
+ V(r)\bigg]R^{\lambda_{\upsilon}}_{n}(r) =
E_{n\upsilon}R^{\lambda_{\upsilon}}_{n}(r), \label{6DRE}
\end{equation}
where we have assumed $m=m_{p}=m_{n}$ (not to be confused with the
number of Jacobi quasiparticles).

For harmonic oscillator potential $V(r)=\frac{1}{2}Cr^{2}$, the
energy spectrum is that of six-dimensional oscillator $E_{N} =
\epsilon (N+\frac{6}{2})$ with $N = 0, 1, 2, \ldots$ and $\epsilon =
\sqrt{C/m}$. This case corresponds to the $U(5)$ dynamical symmetry
limit of the BM model or the IBM.

An $r$-rigid WJ-type model assumes, in addition, that radial
coordinate $r$ is frozen at some non-zero value $r_{0}$. Then, the
radial degree of freedom can be suppressed and the Hamiltonian in
(\ref{6DRE}) reduces to
\begin{equation}
H_{6DWJ} = \frac{\hbar^{2}}{2mr^{2}_{0}} \Lambda^{2}
\label{6DWJ-Ham}
\end{equation}
and its eigenvalues  determine the energies which now are not
equidistant and are given by
\begin{equation}
E_{\upsilon} = \frac{\hbar^{2}}{2mr^{2}_{0}} \upsilon(\upsilon+4)
\equiv A\upsilon(\upsilon+4). \label{6DWJen}
\end{equation}
Note that, in contrast to Eq.(\ref{WJen}), this expression produces
a characteristic ratio $E_{4_{1}}/E_{2_{1}} \simeq 2.67$ of the
ground state band energies, for which $L=\upsilon$ (left diagonal of
Table \ref{SUpn3BS} with $(\lambda,\mu)=(k,0)$, $k=0, 2, 4,
\ldots$). Notice also that the energies (\ref{6DWJen}) are of
kinetic origin and having increasing seniority $\upsilon$.

Usually, the potential energy $V(r)$ is not invariant under
six-dimentional rotations but only under rotations in three
dimensions. The latter means that the potential energy breaks the
$SO(6)$ symmetry to $SO(3)$. Then one could consider the following
algebraic Hamiltonian
\begin{equation}
H = A\Lambda^{2} + BC_{2}[SU_{pn}(3)] + aL^{2}, \label{fullSO6-H}
\end{equation}
which for $B=a=0$ corresponds to the Hamiltonian (\ref{6DWJ-Ham}).
Note that, due to the mutual complementarity given by
Eq.(\ref{O6SU3O2rel}), we can equivalently make use of the $SO(2)$
Casimir operator $M^{2}$ instead of $C_{2}[SU_{pn}(3)]$. We want to
point out that the states (\ref{SU11xSO6wf}) defined by the irreps
of the subgroup chain (\ref{Sp2RxO6-DS}) diagonalize a more general
$r$-soft Wilets-Jean-like model Hamiltonian of the form
\cite{sp2rxso6}:
\begin{align}
H = \
&H\Big(S^{(\lambda_{\upsilon})}_{0},S^{(\lambda_{\upsilon})}_{+},S^{(\lambda_{\upsilon})}_{-}\Big)
+
V(r) \notag\\
&+ A\Lambda^{2} + BC_{2}[SU_{pn}(3)] + aC_{2}[SO(3)],
\label{rsoft-H}
\end{align}
which has an $SU(1,1)$ dynamical group spanned by the generators
$\{S^{(\lambda_{\upsilon})}_{0},S^{(\lambda_{\upsilon})}_{+},S^{(\lambda_{\upsilon})}_{-}\}$.
The latter largely simplifies the diagonalization.

Finally we note also that there is a prolate-oblate degeneracy
related with the conjugate $SU_{pn}(3)$ multiplets $(\lambda,\mu)$
and $(\mu,\lambda)$ contained within the corresponding $SO(6)$
irreducible representations (cf. Table \ref{SUpn3BS}).

\section{Conclusions}

In the present paper, we consider another shell-model coupling
scheme of the PNSM defined through the reduction of the
direct-product group $SU(1,1) \otimes SO(6)$, which acts in the
configuration subspace $\mathbb{R}^{6} \subset \mathbb{R}^{21}$ that
is related to the combined proton-neutron excitations. It is
demonstrated that this subspace is closely related to the
configuration space of the generalized BM model, in which the
monopole degrees of freedom are included together with the
quadrupole ones. This in turn allows to formulate the shell-model
many-particle counterparts of the two exactly solvable limits of the
Bohr-Mottelson model, namely the $\gamma$-unstable and (soft-)rotor
models because the group $SO(6)$ acting in $\mathbb{R}^{6}$ contains
a $SU(3)$ subgroup which irreps could be readily mixed by a more
realistic Hamiltonian. This is a significant result of the
microscopic theory of collective motion in atomic nuclei. In this
respect we recall that a microscopic version of the BM model that is
augmented by the intrinsic vortex-spin degrees of freedom and
compatible with the composite many-nucleon structure of the nucleus
is provided by the (one-component) symplectic model $Sp(6,R)$, which
is sometimes called microscopic collective model. The $Sp(6,R)$
model, however, does not contain $O(5)$ or $O(6)$ structures, which
could allow to associate it with the $\beta$-rigid or $\beta$-soft
but $\gamma$-unstable type dynamics of the Wilets-Jean model. It
contains the Elliott's $SU(3)$ and Ui's rigid-rotor $ROT(3)$ models,
both of which can be associated only with the rotor-model limit of
the Bohr-Mottelson model. From the other side, in contrast to the
original rigid rotor model, the rotational dynamics in the $Sp(6,R)$
model due to the intrinsic vortex-spin degrees of freedom span the
continuous range from irrotational to rigid flows. As a result the
rotational dynamics in the one-component symplectic model possesses
a good $SU(3)$ or $ROT(3)$ dynamical or quasi-dynamical symmetry.
The same type of rotational dynamics can be obtained for the proton,
neutron or combined proton-neutron system in the framework of the
PNSM with $Sp(12,R)$ dynamical group. The present paper shows
further that when the combined proton-neutron dynamics is restricted
to the $\mathbb{R}^{6} \subset \mathbb{R}^{21}$ subspace, spanned by
the six components $Q_{ij}(p,n)$, it can be associated with the
quadrupole-monopole dynamics of the generalized Bohr-Mottelson model
spanning another important class of $\gamma$-unstable collective
models of Wilets-Jean type. The many-particle counterparts of the WJ
and rotor models obtained in the present work are endowed with the
microscopic shell-model wave functions and closely parallel the
relationship between the original BM submodels. In this way an
embedding of the generalized Bohr-Mottelson model in the microscopic
shell-model theory of the nucleus is obtained.

Eqs.(\ref{O6-DS}) or (\ref{Sp2RxO6-DS}) actually introduce another
shell-model coupling scheme within the microscopic proton-neutron
symplectic-based shell-model approach. In principle, this coupling
scheme provides an alternative basis for shell-model diagonalization
of an arbitrary collective Hamiltonian, which could also be
expressed as a polynomial in the many-particle position and momentum
Jacobi coordinates $x_{is}(\alpha)$ and $p_{is}(\alpha)$ in a manner
similar to the ACM. Additionally, any Bohr-Mottelson Hamiltonian of
the form $H = - \frac{\hbar^{2}}{2B}\nabla^{2} + V(\beta,\gamma)$
(\ref{HBohr}) immediately defines a microscopic shell-model
Hamiltonian in which the operator $- \frac{\hbar^{2}}{2B}\nabla^{2}$
is replaced by the many-particle kinetic energy (\ref{Sp12d}), and
$V(\beta,\gamma)$ can be expressed in terms of the microscopic
quadrupole moment operators (\ref{Sp12a}) since $[Q \times Q]^{(0)}
\sim \beta^{2}$ and $[Q \times Q \times Q]^{(0)} \sim
\beta^{3}cos3\gamma$. The difference between the present approach
and the ACM is in the irreducible collective subspaces in which the
model Hamiltonians act. For the microscopic models, like the PNSM,
the state space is defined by allowed $O(A-1)$ (or complementary to
it $Sp(12,R)$) irreducible representations $\omega$ that are
consistent with the Pauli principle, whereas for the
phenomenological models the state space in which the collective
Hamiltonians act is defined by the $O(A-1)$-scalar subspace of the
many-particle Hilbert spaces with $\omega = (0)$. The specific
structure of this violated permutational symmetry space
$\mathbb{H}^{\omega=(0)}$ is that it gives a "deep freezing" of the
microscopic collective features of the used Hamiltonians and make
them similar to those in the Bohr-Mottelson theory, associated with
the irrotational-flow collective dynamics. In this way the results
obtained in the present paper provide us with a fully microscopic
proton-neutron symplectic-based shell-model approach to the
generalized quadrupole-monopole Bohr-Mottelson dynamics that covers
all the range from rigid to irrotational flows. The combined
proton-neutron dynamics in $\mathbb{R}^{6} \subset \mathbb{R}^{21}$
is governed by the microscopic shell-model intrinsic structure of
the symplectic bandhead which defines the Pauli allowed $SO(6)$, and
hence $SU(3)$, subrepresentations. The original Wilets-Jean-type
dynamics of the BM model is recovered for the case of closed-shell
nuclei, for which the symplectic bandhead structure is trivially
reduced to the scalar or equivalent to it representation.

The present many-particle shell-model counterpart of the
Bohr-Mottelson model, represented by the $SU(1,1) \otimes SO(6)$
limit of the PNSM, could be used for a rough and fast evaluation of
different collective observables in the exact limit of the present
approach. This will allow to perform an everyday analysis of the
experimental data with energies and transition probabilities (with
the use of an effective charge) that are very close to the
experimental values. Of course, then, the mixed representation
symplectic-based shell-model calculations with no effective charge
could be performed. All this, as well as the computational technique
required for performing such detailed PNSM shell-model calculations,
including some simple illustrative examples of the physics presented
here is given in Ref.\cite{sp2rxso6}.

\end{document}